# Light-Emitting Diodes with Micrometer-Thick Perovskite Charge Transport Layers

**Sang-Hyun Chin**[1,2]

1. *SKKU Advanced Institute of Nanotechnology (SAINT), Sungkyunkwan University, Suwon 16419, Republic of Korea.*
2. *Department of Physics, Yonsei University, Seoul 03722, Republic of Korea.*

Correspondence to: Sang-Hyun Chin, Department of Physics, Yonsei University, Seoul 03722, Republic of Korea. E-mail: sanghyunchin@yonsei.ac.kr; ORCID: https://orcid.org/0000-0002-0963-4777.

## Abstract

Over the past few decades, thin-film optoelectronic devices have shown significant advancements. Light-emitting diodes (LEDs) based on organic materials, polymers, quantum dots, as well as metal halide perovskites have achieved remarkable efficiencies and long lifetimes, making them ideal for applications in full-color displays and solid-state lighting. These devices typically feature a layered structure, with the light-emitting layer positioned between charge transport layers and two electrodes. This perspective reviews recent progress in LEDs utilizing perovskite charge transport layers and suggests potential pathways for further development in this field.

## INTRODUCTION

Halide perovskite material group has been one of the most competitive candidates in optoelectronic applications such as solar cells and light-emitting diodes (LEDs).[1–3] Owing to the enormous effort from the researchers in the field, those are looking forward to the commercialization.[4,5] This glory is mainly stems from several properties that stand out, such as high charge diffusion length, high luminance or sharp emission spectra, however the high conductivity of halide perovskites are likely related to the most of device applications.[2,6–9] For instance, solar cells with thick perovskite light absorbing layers ensures high photocurrent, maintaining conductivity due to the decent charge transporting nature of perovskites. Based on this point, the high conductivity of perovskites could be still exploited properly when these thin films are used as charge transport layers. Since the first application of perovskite as the charge transport layers in organic light-emitting diodes (OLEDs) by Tian et al. in 2016, there are several more reports have been addressed, however, the reason to continue this research is still quite vague.[10] Hence, to advance the development of this field further and intrigue researchers working on perovskite optoelectronics, the prospects regarding perovskite charge transport layers is assessed, including progress, opportunities, and discuss several practical research paths for the future.

## PEROVSKITE CHARGE TRANSPORT LAYERS

### The Rise of Perovskite Charge Transport Layers

General studies with halide perovskites have been done with non-transparent perovskites due to the use in light-absorbing, mainly iodide perovskites for solar cells or visible-ray-emitting perovskites for LEDs. However, the band gap of halide perovskites could be easily modified by exchange and realized as transparent thin films.[2,11] For this reason, the first form of perovskite charge transport layer is realized with a chloride perovskite, CH3NH3PbCl3 (MAPbCl3), which is transparent in visible region.[10]

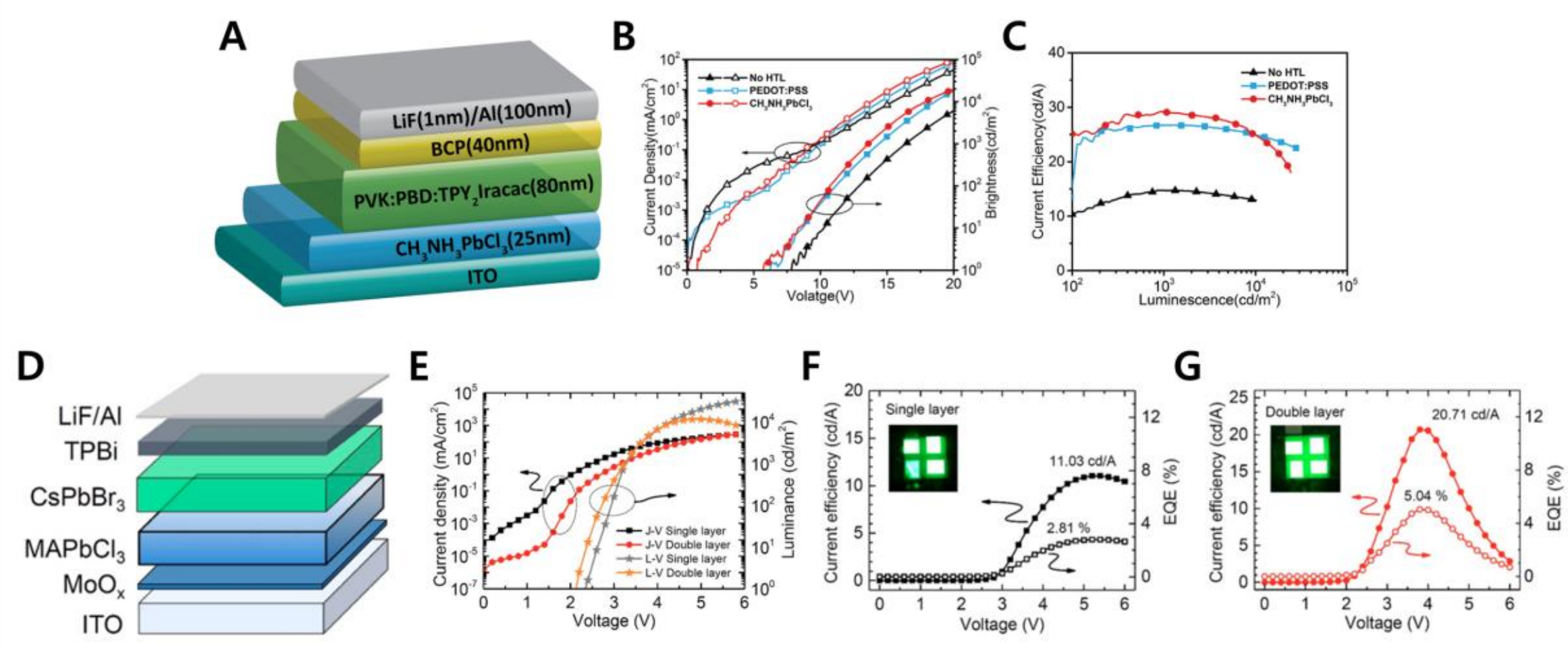

**Figure 1**. First halide perovskite-based charge transport layers in LEDs. A: Structure of OLEDs consist of CH3NH3PbCl3 (MAPbCl3) hole transport layers. B: Current density-voltage-luminance (J-V-L) results comparing the use of Poly(3,4-ethylenedioxythiophene)-poly(styrene-sulfonate) (PEDOT:PSS) and MAPbCl3 hole transport layers. C: Device efficiency as a function of luminance as the use of PEDOT:PSS and MAPbCl3 hole transport layers. These figures are quoted with permission from Tian et al.[10] D: PeLEDs consist of MAPbCl3 hole transport layers. E: J-V-L curves of Single-layered and Double-layered PeLEDs. F: Device efficiency of Single-layered PeLEDs as a function of voltage. G: Device efficiency of Double-layered PeLEDs as a function of voltage. These figures are quoted with permission from Kang et al.[12]

As shown in Figure 1A, MAPbCl3 is employed to form hole transport layer in solution-processed OLEDs with structure of indium tin oxide (ITO) electrodes/perovskite hole transport layer/ Bis[5-methyl-2-(2-pyridinyl-N)phenyl-C](2,4-pentanedionato-O2,O4) iridium (III) (TPY2Ir(acac)) triplet emitter doped in 2-(4-Biphenyl)-5-(4-tert-butylphenyl)-134-oxadiazole (PBD) and Poly(9-vinylcarbazole) (PVK) exciplex forming host/Bathocuproine (BCP) electron transport layer/LiF-aluminum electrodes. Compared to the hole transport layer (HTL)-free devices and PEDOT:PSS-based devices, MAPbCl3 HTL-based LEDs demonstrate lower leakage current in low-operating voltage region yet exhibits higher current and following higher luminance after turn-on, as shown in Figure 1B. As a result of this trend, HTL-free, Poly(3,4-ethylenedioxythiophene)-poly(styrene-sulfonate) (PEDOT:PSS)-based, and perovskite-based OLEDs show current efficiencies of 14.7, 26.6, and 29.2 cd/A (Figure 1C). They claim the improved device performance is plausibly attributed to the suitable energy levels, and efficient hole injection and transport from ITO to MAPbCl3 and light-emitting layer. Interestingly, the gap between the work function of ITO (4.7 eV) to the valance band maximum of MAPbCl3 (5.8 eV) is not negligible, yet the hole injection is still efficient. This might stem from the ionic property of general metal halide perovskites and possibly the mobile ions assist the hole injection against the mismatched band alignment. [2,13,14]

After this first demonstration of perovskite HTL-based OLEDs, Kang et al. reported the same format of study but with perovskite light-emitters onto perovskite HTLs.[12]

Sequentially, ITO electrodes/molybdenum oxide (MoOx) interlayer with dipole/MAPbCl3 hole transport layer/CsPbBr3 light-emitting layer/2,2',2''-(1,3,5-Benzinetriyl)-tris(1-phenyl-1-H-benzimidazole) (TPBi)/LiF-Al electrodes are deposited for this work as shown in Figure 1D. It is noteworthy that solution-processed 3D CsPbBr3 perovskite onto 3D

MAPbCl3 perovskite does not show halide intermixing which probably results in blue emission.[14] As a result, these MAPbCl3 perovskite/CsPbBr3 perovskite double-layered LEDs exhibit less leakage current and lower turn-on voltage in J-V-L results in Figure 1E possibly due to the better morphology from enhanced crystal seeding and blocking of electrons. In addition, improved device efficiency by including MAPbCl3 perovskite HTLs could be confirmed when the "Single-layered" PeLEDs (Figure 1F) and "Double-layered" PeLEDs (Figure 1G) are compared. This work casts a very meaningful message, which is underlying perovskite HTL can assist seeding of following perovskite light-emitting layer and passivates interfacial defects.

## Micrometer-Thick Perovskite Charge Transport Layers

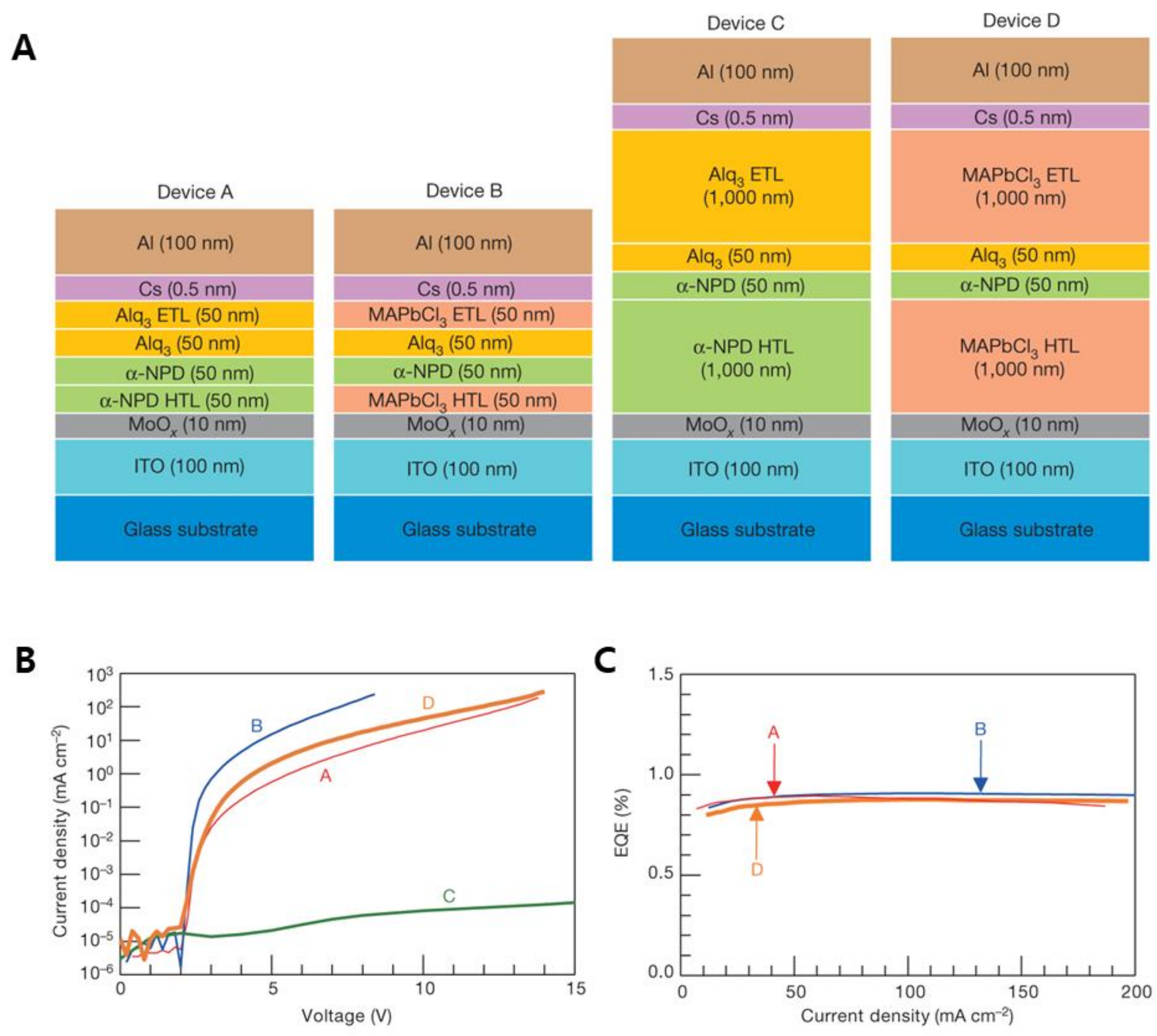

**Figure 2**. OLEDs with micrometer-thick perovskite hole and electron transport layers. A: Device structures of OLEDs with and without perovskite transport layers; Device A-D. B:

Current density-voltage curves of Device A-D. C: External quantum efficiency (EQE) as a function of Device A, B, and D. These figures are quoted with permission from Matsushima et al.[15]

In 2019, Matsushima et al., reported that this perovskite charge transport layers in LEDs could be thick as several micrometers due to the extremely high mobility of perovskites.[15] Firstly they confirmed high mobility; the hole mobilities of 0.9 cm2 V−1 s−1 (for L = 1,000 nm), 1.1 cm2 V−1 s−1 (for L = 2,000 nm) and 1.3 cm2 V−1 s−1 (for L = 3,000 nm) and electron mobilities of 2.1 cm2 V−1 s−1 (for L = 1,000 nm), 2.5 cm2 V−1 s−1 (for L = 2,000 nm), and 2.9 cm2 V−1 s−1 (for L = 3,000 nm), which are nearly independent of the MAPbCl3 thickness. The main purpose of this work is to make tris(8−hydroxyquinoline) aluminum (Alq3) electroluminescent in several fully evaporated OLED structures shown in Figure 2A, using perovskite and N,N'-di(1-naphthyl)-N,N'-diphenyl-(1,1'-biphenyl)-4,4'-diamine (α-NPD) for hole transport. For the electron transport counterpart, perovskite and Alq3 itself are used. The point is that Device C with the micrometer thick organic charge transport layers shows extremely poor current flow while the OLED with micrometer scale perovskite still exhibits decent current as shown in Figure 2B. Interestingly, even the external quantum efficiency of thick perovskite-based OLED has decreased negligible amount, and it even surpasses that of OLED with thin organic charge transport layers (Figure 2C). This finding that extremely thick charge transport layer solves a huge difficulty in large scale mass production of OLED industry. Covering a large substrate uniformly is already a bit hard and variations in thickness cause the formation of shunting paths between electrodes, thereby lowering device production yield. To overcome this issue, thicker organic transport layers are desirable because they can cover particles and residue on substrates, but increasing their thickness increases the driving voltage as demonstrated in their work. Thus, when the highly transparent and conductive perovskites are deposited in micrometer scale, there must be a huge breakthrough in this issue.

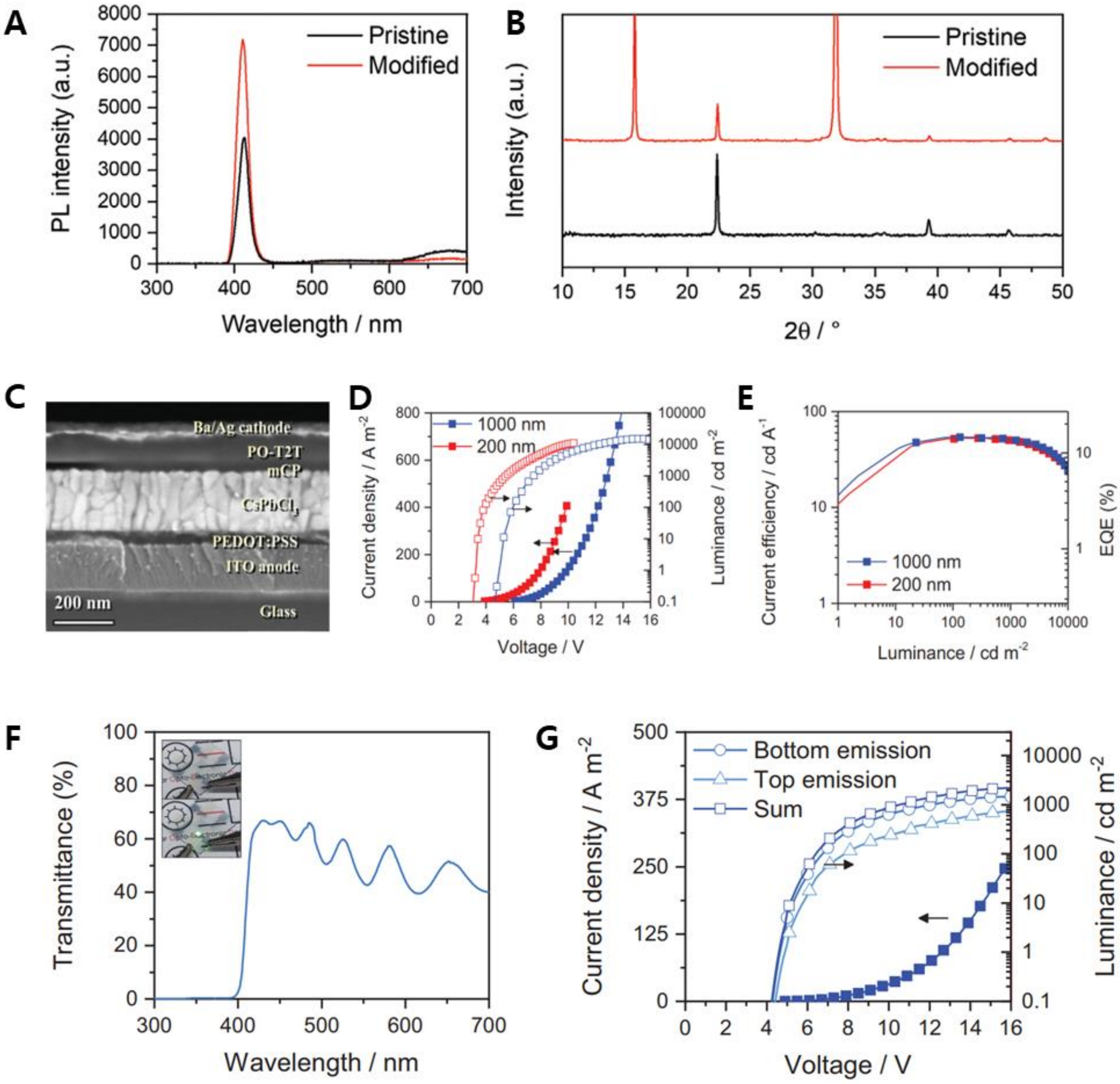

**Figure 3**. Transparent OLEDs with sub-nanometer thick triplet light-emitters and micrometer-thick perovskite hole transport layers. A: Photoluminescent intensity of pristine and CsCl-modified CsPbCl3 thin films. B: X-ray diffraction patterns of pristine and CsCl-modified CsPbCl3 thin films. C: Structure of OLEDs includes perovskite HTLs confirmed by cross-section scanning electron microscope imaging. D: J-V-L curves of OLED with 200 nm-thick and 1000 nm-think perovskite HTLs. E: Device efficiency as a function luminance of OLED with 200 nm-thick and 1000 nm-think perovskite HTLs. F: Transmittance of OLED stacks (inset: operating transparent OLED). G: J-V-L curve of transparent OLEDs obtained from both sides. These figures are quoted with permission from Forzatti et al.[16]

Forzatti et al., recently published an interesting report comprising micrometer-thick CsPbCl3 and ultrathin sub-nanometer-thick iridium complex which is a triplet emitter.[16]

Deposition of this ultrathin triplet emitter has been employed to get rid of the risk of triplet-triplet annihilation and following decrease of luminescence.[17] Yet, this causes shunting paths and other issues that Matsushima et al. had to solve as mentioned above. The work by Forzatti et al. properly employed co-evaporated perovskite HTLs for this ultrathin emitter-based OLEDs. In addition, an essential treatment with insight is performed. During the co-evaporation process, different portions of the intended amount get deposited onto the substrates due to varying sticking coefficients.[18] Mostly, lead halides, PbCl2 in this case, have better chance to get stick on, hence they evaporated several nanometers of CsCl prior to the deposition of perovskites. The PL spectra of pristine CsPbCl3 and modified CsPbCl3 films were measured as shown in Figure 3A, exhibiting that the PL peak of modified film is improved by almost two times, proving that the surface defects are reduced by the excess CsCl on interfaces. In addition, the XRD pattern (Figure 3B) of modified perovskite layers shows two extra peaks at 15.8 ° and 31.8 °, which can be indexed as (100) and (200), besides the peak at 22.3 ° (101) present in the pristine film. This difference indicates that the seeding CsCl layer exerts a significant influence on the subsequent crystal growth of the deposited CsPbCl3 layer. Specifically, it suggests that the presence of the CsCl layer induces preferential crystal growth along the axis denoted by the first crystallographic index.[19–21] Based on this improved perovskite HTL, they evaporated 0.03 nm Bis(2-phenyl-pyridine) acetylacetonate iridium (III) (Irppy2(acac)) OLEDs exploiting interface exciplex pair; 1,3-Bis(N-carbazolyl)benzene (mCP) and 2,4,6-Tris[3-(diphenyl-phosphinyl) phenyl]-1,3,5-triazine (PO-T2T) as shown in the cross-section image obtained by scanning electron microscopy (Figure 3C). The device performance is provided in Figure 3D, which describes J-V-L characteristics of OLEDs with 200 nm-thick perovskite and 1000 nm-thick perovskite HTLs. Due to a bit higher resistance from thicker perovskite layer, the turn-on voltage became higher, however the device efficiency, of both type of OLEDs as a function of luminance, are near 55 cd/A with negligible differences (Figure 3E).

Also, pulsed laser deposited ITO electrodes are examined completing the device stack as: ITO (150 nm)/PEDOT:PSS (30 nm)/modified perovskite HTL (1000 nm)/mCP (20 nm)/Ir(ppy)2acac (0.03 nm)/PO-T2T (60 nm)/Ba (5 nm)/Ag (5 nm)/ITO (135 nm). A transmission between 40% and 70% was demonstrated across the visible region spanning from 410 to 700 nm, with an average visible transmittance (AVT) value of 50.2% in Figure 3F. The J-V-L characteristics of the devices were considered, with the luminance measured simultaneously from the top (Ba/Ag/ITO) and the bottom (glass/ITO) side in Figure 3G. The peak luminance of the semitransparent device reaches a value of 1453 cd m-2 when measured from the bottom side and 829 cd m-2 from the top side. While the obtained values are still high enough to guarantee a very strong on/off contrast in conjunction with the high transparency, they are significantly lower than what is obtained for the opaque device with Ag electrodes (over 15 000 cd m-2). This is plausibly due to the lower charge injection and current density and a lower peak efficiency of 41.3 cd A-1 considering both top and bottom emission together, which in turn might be the result of lower efficiency in light extraction.

## OUTLOOK AND PERSPECTIVE

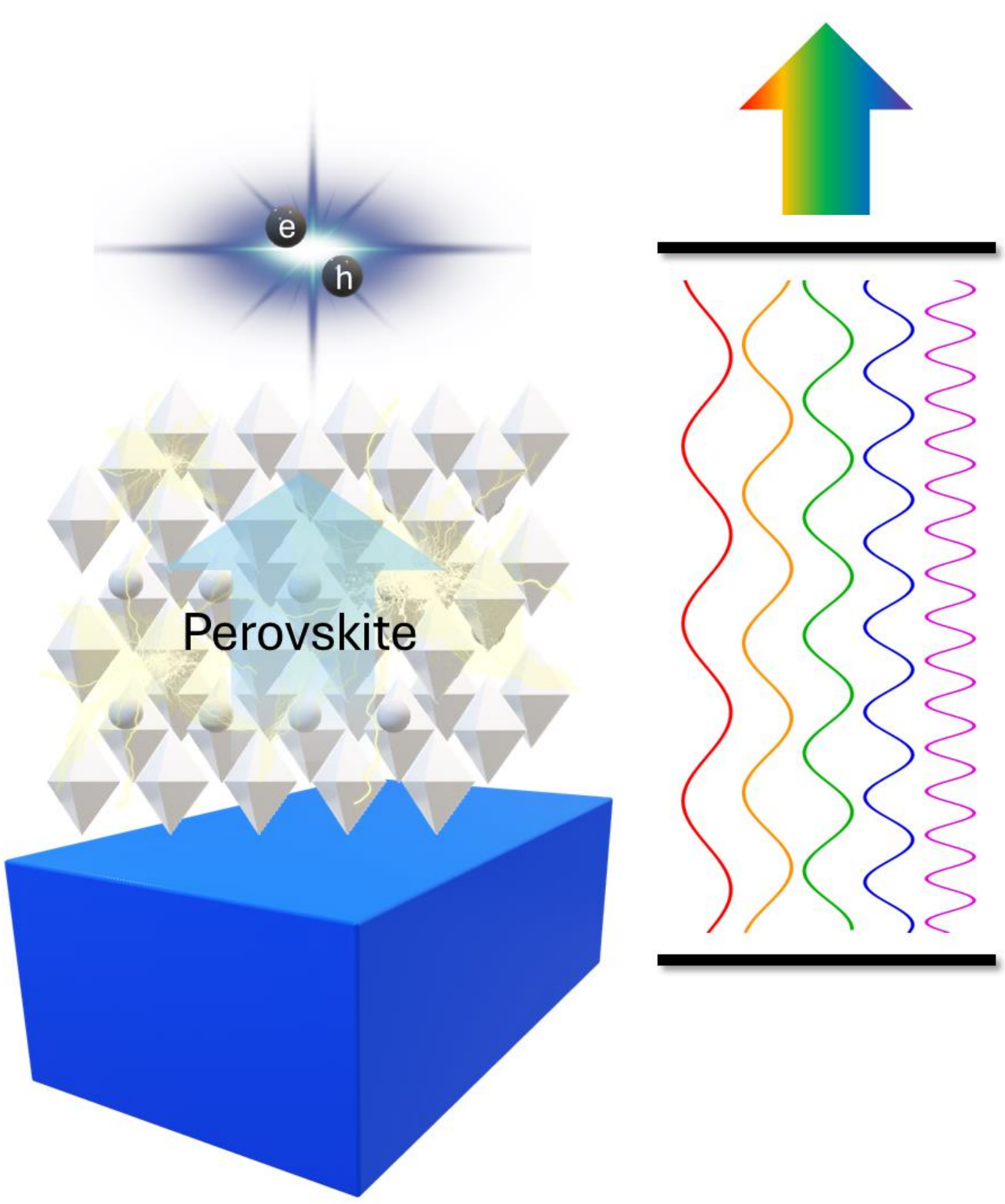

**Figure 4**. Perovskite-based charge-transporting thickness modulators in LEDs as cavities.

As explained above, the perovskite charge transport technology in LED has been developed remarkably using both solution process and thermal evaporation process. Until now, there is a lack of research on micrometer thick perovskite charge transport layers possibly due to the limited management of defects.[22] Possibly this issue can be addressed with understanding complexation of precursors and Lewis-base or proper aging.[23,24] Several examples introduced above employ thermally evaporated micrometer-thick transport layers, but the aim is heavily focused on completing specific 1 micrometer thick. These highly conductive perovskite polycrystalline layers show nearly thickness-independent characteristics in device efficiency. This implies that thickness of transparent perovskite charge transport layer can be optimized from an optical point of view in cavity system as depicted in Figure 4. By a careful management of thickness, the perovskites and transparent hybrid materials can be an optical modulator in cavity system, which can be applied to lasing system or photoconductor in tandem photodetectors as well as tandem LED systems.[25–31] After a full exploitation of these points from the strengths of perovskites, the technique introduced in this perspective might open completely new horizon for researchers giving enormous effort to metal halide perovskites.

## Availability of data and materials

Not applicable.

## Financial support and sponsorship

This research was supported by Sungkyunkwan University and the BK21 FOUR (Graduate School Innovation) funded by the Ministry of Education (MOE, Korea) and National Research Foundation of Korea (NRF).

## Conflicts of interest

The author declared that there are no conflicts of interest.

## REFERENCES

[1] M. Grätzel, Acc. Chem. Res. 2017, 50, 487.

[2] A. Fakharuddin, M. K. Gangishetty, M. Abdi-Jalebi, S.-H. Chin, Abd. R. bin Mohd Yusoff, D. N. Congreve, W. Tress, F. Deschler, M. Vasilopoulou, H. J. Bolink, Nat. Electron. 2022, 5, 203.

[3] S.-H. Chin, Journal of Optics and Photonics Research 2024, 2 (3), 119-127. DOI: 10.47852/bonviewjopr42022936.

[4] J. Luo, J. Li, L. Grater, R. Guo, A. R. bin Mohd Yusoff, E. Sargent, J. Tang, Nat. Rev. Mater. 2024, DOI: 10.1038/s41578-024-00651-8.

[5] Z. Xu, S.-H. Chin, B.-I. Park, Y. Meng, S. Kim, S. Han, Y. Li, D.-H. Kim, B.-S. Kim, J.-W. Lee, S.-H. Bae, Next Materials 2024, 3, 100103.

[6] B. Turedi, M. N. Lintangpradipto, O. J. Sandberg, A. Yazmaciyan, G. J. Matt, A. Y. Alsalloum, K. Almasabi, K. antyn Sakhatskyi, S. Yakunin, X. Zheng, R. Naphade, S. Nematulloev, V. Yeddu, D. Baran, A. Armin, M. I. Saidaminov, M. V Kovalenko, O. F. Mohammed, O. M. Bakr, Advanced Materials 2022, n/a, 2202390.

[7] M. J. P. Alcocer, T. Leijtens, L. M. Herz, A. Petrozza, H. J. Snaith, Science (1979). 2013, 342, 341.

[8] S. H. Chin, J. W. Choi, H. C. Woo, J. H. Kim, H. S. Lee, C. L. Lee, Nanoscale 2019, 11, 5861.

[9] H. C. Woo, J. W. Choi, J. Shin, S. H. Chin, M. H. Ann, C. L. Lee, Journal of Physical Chemistry Letters 2018, 9, 4066.

[10] Y. Tian, Y. Ling, Y. Shu, C. Zhou, T. Besara, T. Siegrist, H. Gao, B. Ma, Adv. Electron. Mater. 2016, 2, DOI: 10.1002/aelm.201600165.

[11] K. Young-Hoon, C. Himchan, L. Tae-Woo, Proceedings of the National Academy of Sciences 2016, 113, 11694.

[12] D. H. Kang, S. G. Kim, Y. C. Kim, I. T. Han, H. J. Jang, J. Y. Lee, N. G. Park, ACS Energy Lett. 2020, 5, 2191.

[13] P. Mao, X. Shan, H. Li, M. Davis, Q. Pei, Z. Yu, ACS Appl. Electron. Mater. 2021, DOI: 10.1021/acsaelm.1c01201.

[14] S. Chin, L. Mardegan, F. Palazon, M. Sessolo, H. J. Bolink, 2022, DOI: 10.1021/acsphotonics.2c00604.

[15] T. Matsushima, F. Bencheikh, T. Komino, M. R. Leyden, A. S. D. Sandanayaka, C. Qin, C. Adachi, Nature 2019, 572, 502.

[16] M. Forzatti, S. H. Chin, M. A. Hernández-Fenollosa, M. Sessolo, D. Tordera, H. J. Bolink, Adv. Opt. Mater. 2024, DOI: 10.1002/adom.202401061.

[17] Y. Miao, M. Yin, iScience 2022, 25, 103804.

[18] B. S. Kim, L. Gil-Escrig, M. Sessolo, H. J. Bolink, Journal of Physical Chemistry Letters

2020, 11, 6852.

[19] C. Ma, † Felix, T. Eickemeyer, † Sun-Ho Lee, D.-H. Kang, S. J. Kwon, M. Grätzel, N.-G. Park, Unveiling Facet-Dependent Degradation and Facet Engineering for Stable Perovskite Solar Cells, n.d.

[20] C. Ma, M. C. Kang, S. H. Lee, Y. Zhang, D. H. Kang, W. Yang, P. Zhao, S. W. Kim, S. J. Kwon, C. W. Yang, Y. Yang, N. G. Park, J. Am. Chem. Soc. 2023, 145, 24349.

[21] C. Ma, M. C. Kang, S. H. Lee, S. J. Kwon, H. W. Cha, C. W. Yang, N. G. Park, Joule 2022, 6, 2626.

[22] Y. Zhou, M. Li, W. Yin, Q. Zeng, Y. Zhao, X. Zhang, Small 2026, DOI: 10.1002/smll.202512877.

[23] J. Park, J. Kim, H. S. Yun, M. J. Paik, E. Noh, H. J. Mun, M. G. Kim, T. J. Shin, S. Il Seok, Nature 2023, 616, 724.

[24] S. H. Chin, J. W. Choi, Z. Hu, L. Mardegan, M. Sessolo, H. J. Bolink, J. Mater. Chem. C Mater. 2020, 8, 15996.

[25] S. H. Chin, D. Cortecchia, M. Forzatti, C. S. Wu, A. L. Alvarado-Leaños, G. Folpini, A. Treglia, I. A. Kalluvila Justin, A. Paliwal, C. Cho, C. Roldán-Carmona, M. Sessolo, A. Petrozza, H. J. Bolink, Adv. Opt. Mater. 2024, DOI: 10.1002/adom.202302701.

[26] S.-H. Chin, Discover Applied Sciences 2024, 6, 396.

[27] L. Martínez-Goyeneche, L. Gil-Escrig, I. Susic, D. Tordera, H. J. Bolink, M. Sessolo, Adv. Opt. Mater. 2022, 10, DOI: 10.1002/adom.202201047.

[28] L. Kong, Y. Luo, Q. Wu, X. Xiao, Y. Wang, G. Chen, J. Zhang, K. Wang, W. C. H. Choy, Y. B. Zhao, H. Li, T. Chiba, J. Kido, X. Yang, Light Sci. Appl. 2024, 13, DOI: 10.1038/s41377-024-01500-7.

[29] M. Zhu, S. Q. Sun, W. He, Y. L. Xu, Q. Sun, Y. M. Xie, M. K. Fung, S. T. Lee, J. Mater. Chem. C Mater. 2024, 12, 2623.

[30] H. D. Lee, S. J. Woo, S. Kim, J. Kim, H. Zhou, S. J. Han, K. Y. Jang, D. H. Kim, J. Park, S. Yoo, T. W. Lee, Nat. Nanotechnol. 2024, 19, 624.

[31] S. H. Chin, D. Lee, D. Lee, S. Kim, B. Kang, K. Chung, T. Il Kim, J. Yeon, S. H. Lee, S. W. Bae, W. Kim, S. Park, K. Kim, Y. H. Kim, Y. Yi, Advanced Science 2025, , 12, e13328, DOI: 10.1002/advs.202513328.